\documentclass[final,5p,times,twocolumn,authoryear]{elsarticle}

\usepackage{amssymb}
\usepackage[dvipsnames]{xcolor}
\usepackage[colorlinks=true,linkcolor=blue, citecolor=blue, urlcolor=blue]{hyperref}

\journal{Neural Networks}

\begin{document}

\begin{frontmatter}



\title{Reservoir Computing with 3D Nanowire Networks}

\author[label1]{R. K. Daniels}
\author[label1]{J. B. Mallinson}
\author[label2]{Z. E. Heywood}
\author[label2]{P. J. Bones}
\author[label3]{M. D. Arnold}
\author[label1]{S. A. Brown\corref{cor1}}
\affiliation[label1]{organization={The MacDiarmid Institute for Advanced Materials and Nanotechnology, School of Physical and Chemical Sciences, Te Kura Mat\={u}, University of Canterbury, Private Bag 4800, Christchurch 8140, New Zealand}}
\affiliation[label2]{organization={Electrical and Computer Engineering, University of Canterbury, Private Bag 4800, Christchurch 8140, New Zealand}}

\affiliation[label3]{organization={School of Mathematical and Physical Sciences, University of Technology Sydney, PO Box 123
Broadway NSW 2007, Australia}}
\cortext[cor1]{Corresponding author at: The MacDiarmid Institute for Advanced Materials and Nanotechnology, School of Physical and Chemical Sciences, Te Kura Mat\={u}, University of Canterbury, Private Bag 4800, Christchurch 8140, New Zealand \\ \textit{E-mail address:} simon.brown@canterbury.ac.nz}

\begin{abstract}
Networks of nanowires are currently being explored for a range of applications in brain-like (or neuromorphic) computing, and especially in reservoir computing (RC). Fabrication of real-world computing  devices requires that the nanowires are deposited sequentially, leading to stacking of the wires on top of each other. However, most simulations of computational tasks using these systems treat the nanowires as 1D objects lying in a perfectly 2D plane -- the effect of stacking on RC performance has not yet been established. Here we use detailed simulations to compare the performance of perfectly 2D and quasi-3D (stacked) networks of nanowires in two tasks: memory capacity and nonlinear transformation. We also show that our model of the junctions between nanowires is general enough to describe a wide range of memristive networks, and consider the impact of physically realistic electrode configurations on performance. We show that the various networks and configurations have a strikingly similar performance in RC tasks, which is surprising given their radically different topologies. Our results show that networks with an experimentally achievable number of electrodes perform close to the upper bounds acheivable when using the information from every wire. However, we also show important differences, in particular that the quasi-3D networks are more resilient to changes in the input parameters, generalizing better to noisy training data. Since previous literature suggests that topology plays an important role in computing performance, these results may have important implications for future applications of nanowire networks in neuromorphic computing.
\end{abstract}




\end{frontmatter}

\section{Introduction}
\label{intro}
Artificial neural networks (ANNs) underpin machine learning, both in research and in industry applications \citep{goodfellow2016}. However all software implementations of ANNs on conventional machines are subject to the limitations inherent in the underlying von Neumann architecture of the hardware. While CPUs and GPUs allow parallel processing, an inevitable departure from Moore's Law \citep{moore1965,khan2018} has motivated an increased interest in ways to circumvent limitations in fabrication technology, and brain-inspired hardware systems show considerable promise \citep{merolla2014,markovic2020,christensen2021}.
\begin{figure*}[ht!]
	\centering
	\includegraphics[width=\textwidth]{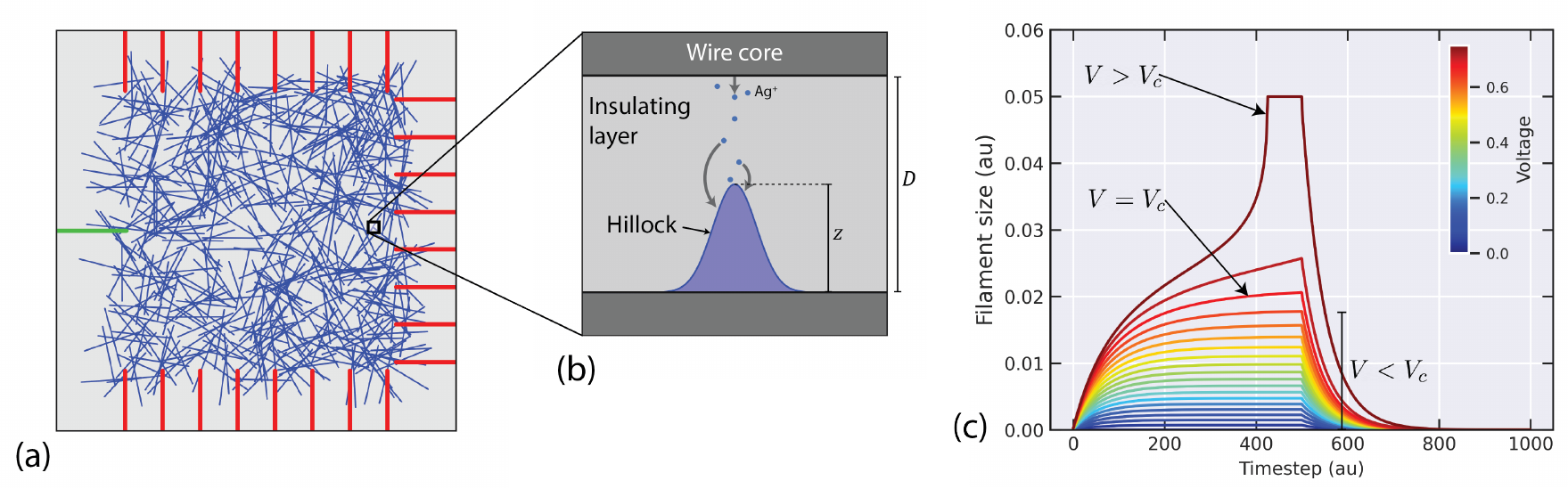}
	\caption{\textit{(a) Example schematic of a 500NWN with a 1 input (green) and 24 output (red) electrode configuration. (b) Illustration of a possible growth mechanism that drives the dynamics within the junction between two silver nanowires - as an electrical potential difference occurs between two contacting wires, silver ions migrate into the junction, forming a filament or hillock. (c) The idenitification of three regimes of junction behaviour - when the voltage is below a critical value $V_c$, the filament grows towards an equilibrium state; when $V_c$ is exceeded, the filament growth undergoes a change in stability, growing rapidly. When the filament grows across the entire width of the gap, a conductive bridge is formed and the junction switches into a low resistance state.}}
	\label{fgr:junctionModel}
\end{figure*}

Reservoir computing (RC) is a neuromorphic approach that was originally conceived as a specialized adaptation of recurrent neural networks (RNN). It was proposed that the input weights and the internal weights in the network should be fixed, leaving only the output weights to be trained with an efficient algorithm such as linear regression \citep{lukovsevivcius2009}, greatly reducing the computational cost of training. However, this prototypical form of RC, called an Echo State Network (ESN) \citep{jaeger2002b}, must still be implemented in software on conventional computer architectures. Hardware implementations of RC have the potential to provide significant performance advantages by circumventing the von Neumann bottleneck and decreasing power consumption. Over the last few years, a significant effort has been invested in novel hardware systems for RC \citep{tanaka2019, nakajima2020}.

To be successful as a reservoir, a system must be able to non-linearly map the inputs so that they become linearly separable in a higher dimensional feature space. An additional requirement is short-term memory such that the current state of the reservoir is influenced by recent past states and not more distant past states. The final requirement is the separation property - the reservoir must be able to distinguish between distinct signals and yet be insensitive to noise such that those signals which are sufficiently similar can still be classified as the same \citep{tanaka2019}.

Here we investigate the potential of nanowire networks (NWNs -- Figure \ref{fgr:junctionModel}a) to act as reservoirs using detailed simulations. Nanowires are typically comprised of either a metal-oxide or metal core and are coated in an insulating material such as a metal-oxide, a chalcogenide electrolyte, or a polymer (for example TiO$_2$ \citep{li2020}, Ag$_2$S \citep{sillin2013}, or PVP \citep{milano2020, diaz2019} respectively). When the wires make contact, they form a metal-insulator-metal (MIM) memristive junction \citep{kuncic2021}, as illustrated in Figure \ref{fgr:junctionModel}b. A memristor is a two terminal electronic device whose resistance responds nonlinearly to changes in input and whose state exhibits memory of previous inputs \citep{strukov2008}.  A voltage bias can be applied across the MIM junction, causing the migration of ions. Depending on the system, an atomic or nanoscale filament begins to form across the junction 
\citep{krishnan2016, yang2020, wang2019}. When this filament forms a complete bridge, the junction experiences a sharp transition from a nonlinear low conductance state to a high conductance Ohmic state. In the low conductance state, which is the focus of the present work, current flow across a junction is via quantum tunnelling. As the filament grows (shrinks), the width of the tunnelling barrier decreases (increases), causing a nonlinear change in conductance for each junction. The coupling of nonlinear junction dynamics to the complex network topology via Kirchhoff's Laws results in rich recurrent dynamics that can be exploited for reservoir computing applications \citep{zhu2020}.

Self-assembled NWNs have previously been investigated and have been shown to exhibit complex brain-like behaviour such as neural avalanching \citep{hochstetter2021,dunham2021}, short- and long-term memory \citep{diaz2020,li2020}, and multi-tasking \citep{loeffler2021}. Demonstrations of neuromorphic behaviour include: winner-takes-all behaviour \citep{manning2018}, higher harmonic generation (HGG) \citep{cohen2012, sillin2013}, and sine wave generation and temporal signal processing \citep{zhu2020b, kuncic2020, fu2020}. Simulations show the promise of NWNs for neuromorphic applications, but experimental demonstrations are thus far very limited \citep{kuncic2021, milano2021, hochstetter2021}.

Simulations of NWNs generally assume the wires to be 1-dimensional objects that interpenetrate each other and hence can lie on a perfectly 2-dimensional plane. However, real nanowires are 3-dimensional objects, and will inevitably become stacked on top of one another during deposition (images of stacking effects in experimental nanowire devices can be seen in \cite{milano2020} and \cite{lee2017}). It has been shown previously that there are significant differences in the topological structure of 2D and quasi-3D (Q3D) NWNs \citep{daniels2021}. This manifests as important differences in network characteristics such as the degree of clustering and average path length between wires. Recent work has also shown that 2D NWNs have a much higher small-world propensity than quasi-3D networks of nanowires \citep{pantone2018,loeffler2020,daniels2021}. It has been suggested that in general, information processing and signal propagation are enhanced in small-world ESN \citep{kawai2019} and coupled oscillator systems \citep{nishikawa2003}. However, it is currently not known how these different network characteristics will affect RC performance.

Most previous simulations of NWN assume that the signals from every wire (node) in the network can be used as outputs for RC. However these signals are in fact inaccessible in real-world experiments, and therefore the performance of realistic networks could be significantly different. In the present work, we present results obtained using a physically realistic electrode configuration (Figure \ref{fgr:junctionModel}a) that reflects the true potential of NWNs for RC applications. We compare the performance of realistic Q3D NWNs with their purely 2D counterparts and find some striking similarities. Specifically, we investigate the performance of the NWNs as physical reservoirs in two tasks: the memory capacity (MC), and the nonlinear transformation (NLT). In addition, we test the separation property by investigating the performance of the NWNs in response to noisy input data. Importantly, we also consider the impact of the number and type of output electrodes that are connected to the NWNs.

\section{Nanowire Networks}
\label{networks}
Here we describe the model of the junction dynamics and the details of the  construction of the NWN.

\subsection{Junction model}
The most basic dynamical element within a NWN is the junction that results from two wires making contact. We model the internal dynamics of the memristive junction as a conductive `hillock' (Figure \ref{fgr:junctionModel}b) that grows through the insulating layer between wires under the influence of an electric field, creating a non-equilibrium structure. Surface energy effects will attempt to reduce the size of this protrusion \citep{wang2019}. The dynamical equation governing the growth of the hillock is then
\begin{equation}
    \frac{\mathrm{d}z}{\mathrm{d}t} = \mu \frac{V}{D - z} - \kappa z \label{eq:junc},
\end{equation}
where $z$ is the height of the hillock and $D$ is the distance between wire cores as in Figure \ref{fgr:junctionModel}b. $V$ is the voltage across the junction, and $\mu$ and $\kappa$ are the parameters which govern growth and relaxation respectively ($\mu = 0.346$nm$^2$V$^{-1}$ and $\kappa = 0.038$s$^{-1}$ -- see discussion below). Eq (\ref{eq:junc}) can then be numerically integrated using the Euler method to obtain a discretized change in hillock height  at each computational timestep. The conductance of each junction is governed by quantum mechanical tunneling, and so is a function of the instantaneous size of each tunnel gap, $D - z$:
\begin{equation}
    G = \alpha e^{-\beta(D - z)}. \label{eq:cond}
\end{equation}
Hence each junction has a nonlinear response to a voltage input. Note that the response of the \textit{network} results from the interaction of all the junctions.

Using Eq (\ref{eq:junc}), we can identify a critical voltage $V_c$. For $V<V_c$, the hillock grows towards a stable equilibrium value. For $V > V_c$, the hillock will grow until $z=D$, forming a conductive bridge between the two wires (Figure \ref{fgr:junctionModel}c). We focus here on the low voltage regime where no hillock  completely bridges a tunnel gap.

We now show that the junction model given in Eq (\ref{eq:junc}) is equivalent to that of an ideal voltage-driven memristor \citep{strukov2008} parameterized using an internal variable $w$ and governed by the dynamical equation
\begin{equation}
	\frac{\mathrm{d}w}{\mathrm{d}t} = \mu\frac{R_{ON}}{D}\frac{V}{R} - \kappa w,
\end{equation}
with the resistance across the gap of size $D$, given by
\begin{equation}
	R = R_{ON}\frac{w}{D} + R_{OFF}\left(1-\frac{w}{D}\right),
\end{equation}
where $R_{ON}$ and $R_{OFF}$ are the resistance across the memristor in the high conductive state and the low conductive state respectively. If we assume \citep{caravelli2019} a fixed ratio of $r = R_{OFF}/R_{ON}$, we can combine Eq (3) and (4) to get
\begin{equation}
	\frac{\mathrm{d}w}{\mathrm{d}t} = r\mu\frac{V}{D-\chi w} - \kappa w,
\end{equation}
where $\chi=(R_{OFF} - R_{ON})/R_{OFF}$. The ratio $r$ is typically on the order of $10^3 - 10^5$ \citep{kuncic2021}, so when $r$ is large, $\chi \to 1$. The resulting equation is then
\begin{equation}
	\frac{\mathrm{d}w}{\mathrm{d}t} = r\mu\frac{V}{D - w} - \kappa w.
\end{equation}
When compared with Eq (\ref{eq:junc}), we can see that they are equivalent up to a factor of $r$ in the first term, which has the effect of renormalizing the relevant range of $\mu$ (note that $r$ is a parameter of the memristor model, but not our simulations). Hence the hillock growth model is equivalent to models of diffusive unipolar memristive junctions operating in a low voltage regime, and the results presented in this paper are applicable to the more general case of nanowire networks with memristive junctions.

We note that other authors have considered alternate regimes and junction models, including bipolar switching models. For example \cite{hochstetter2021} and \cite{loeffler2021} consider a regime where each hillock can extend close to the opposite side of the junction gap, leading to a dramatic increase in conductance which then redistributes the voltage across other junctions. In this regime, conductive bridges can continuously evolve throughout the network, leading to more complex dynamical behaviour. We emphasize that we consider the low voltage regime where the hillocks never completely bridge the junction gaps.

\begin{figure*}[ht!]
	\centering
	\includegraphics[width=\textwidth]{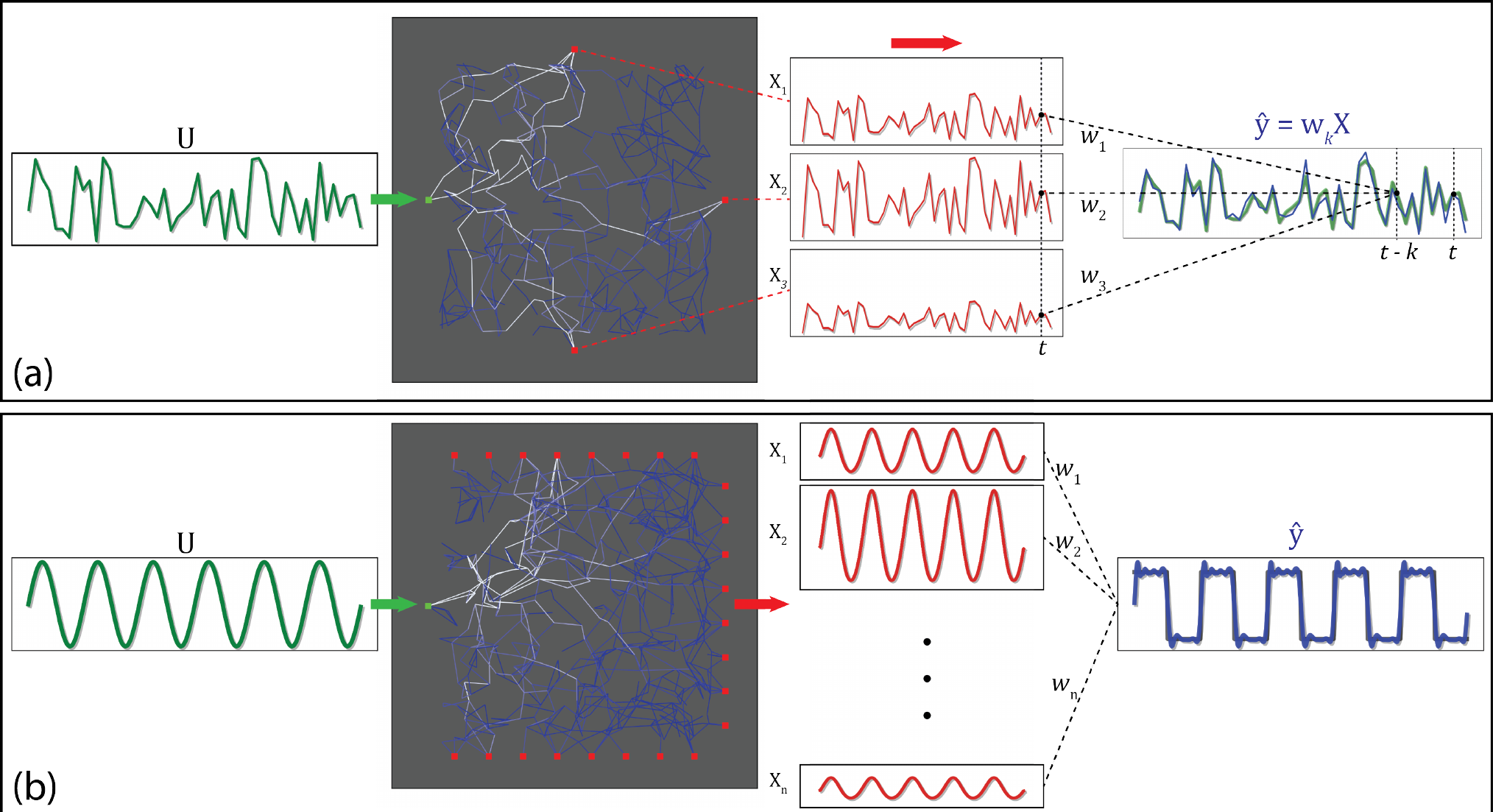}
	\caption{\textit{Schematic of some example networks with different output electrode configurations. Note that the outputs and reconstructed waveforms are examples only, and not meant to be indicative of the real network outputs. (a) The memory capacity calculation shown here with a Q3D 400NWN with 1 input electrode (green square) and 3 output electrodes (red squares). The colored lines within the network represent the current flow across the network (light to dark blue corresponds to a high to low current). The input signal $\mathbf{U}$ (green) is fed into the input electrode and the current $\mathbf{X}_n$ is recorded from each of the $n$ output electrodes (red). The output values at time $t$ are then used to find the predicted output $\mathbf{y}$ (shown by the blue line on the far right) at time $t-k$. (b) The nonlinear transformation process shown here with a Q3D 500NWN with 24 output electrodes. The input waveform $\mathbf{U}$ (green) is fed into the input electrode, and the current is recorded from the outputs (red). Linear regression is then used on the electrode outputs to find the total output $\mathbf{\hat{y}}$ to match the target function $\mathbf{y}$. This general procedure is the same regardless of the selected number of output signals and regardless of whether the output signal is the current at each electrode or the voltage at each node/wire.}}
	\label{fgr:NLTandMC}
\end{figure*}

\subsection{Network simulation}
We constructed two different types of NWNs, 2D and Q3D. The $x-$ and $y-$coordinates of the centres of the wires were chosen randomly from a uniform distribution in the range $[0, \Lambda]$, where $\Lambda$ is the width of the square deposition area. The orientation of the wires with respect to the $x$-axis were drawn randomly from a uniform distribution in the range $[-\pi/2, \pi/2]$. Every point where two wires intercept is considered to be a junction in the network. The connectivity is then stored in an adjacency matrix. 

In the 2D networks, the wires are rigid 1D lines, and hence all lie within the same plane. However, in the Q3D networks, the wires are rigid 3D volumes in space. The positioning of a new wire therefore depends upon those that have been previously deposited. For a new wire, the algorithm determines all potential intercept points, and in the general case, the point of intercept with the highest $z$ value acts as the first contact point and as a pivot around which the new wire must rotate. The center of mass of the new wire then determines the direction of rotation and hence the second contact point. This results in a stacking of the wires in the vertical direction, leading to very different topological properties of the 2D and Q3D networks (for more details, see \cite{daniels2021}).

The input and output electrodes are defined by placing additional nanowires in the deposition area to act as contacts (they are placed before the deposition of the NWN). The input electrode is deposited on the left edge of the deposition area, and the output electrodes are evenly distributed along the top, bottom, and right edges (as in Figure \ref{fgr:junctionModel}a). The `contact' wires are then treated in the same way as all other wires, with additional rows and columns added to the adjacency matrix representation of the network. The input signal is a voltage that is applied to the input electrode, and the output electrodes are grounded. We then solve Kirchhoff's circuit laws for the network, and the current is recorded from each of the output electrodes. All simulations were implemented in Python v3.8.5.

\section{Tasks}
The aim of this paper is to evaluate the potential of the networks for reservoir computing applications and to compare performance for different network topologies. We therefore implement two tasks to test two different capabilities: the memory capacity (MC) which tests the fading memory property of the reservoir, and the nonlinear transformation task (NLT), which tests the ability of the network to map the input onto a higher-dimensional feature space. We perform the tasks using two different methods of reading the network output. In the first, the currents flowing from the output electrodes are used as the readout  (the red electrodes/wires in Figure \ref{fgr:junctionModel}a). In the second, the voltages at every wire (node) provide the output signals (the blue wires in Figure \ref{fgr:junctionModel}a). This second method allows us to exploit information within the network that is normally inaccessible in real experiments due to technological limitations. We expect that these results therefore will provide a theoretical upper bound on the performance. The first approach provides an indication of the practical reliability of real world devices (which necessarily have a limited number of electrodes). For brevity we refer to two different readout methods as E-mode (outputs are electrode currents) and  N-mode (outputs are wire voltages).

We now provide general details of the tasks that are not specific to any particular electrode configuration.

\subsection{Memory Capacity}
The memory capacity task aims to reconstruct a delayed version of the input signal from the measured outputs \citep{jaeger2002}. In other words, the task determines how precisely the reservoir can reconstruct the input of previous time-steps.
\begin{figure*}[!htbp]
	\centering
	\includegraphics[width=\textwidth]{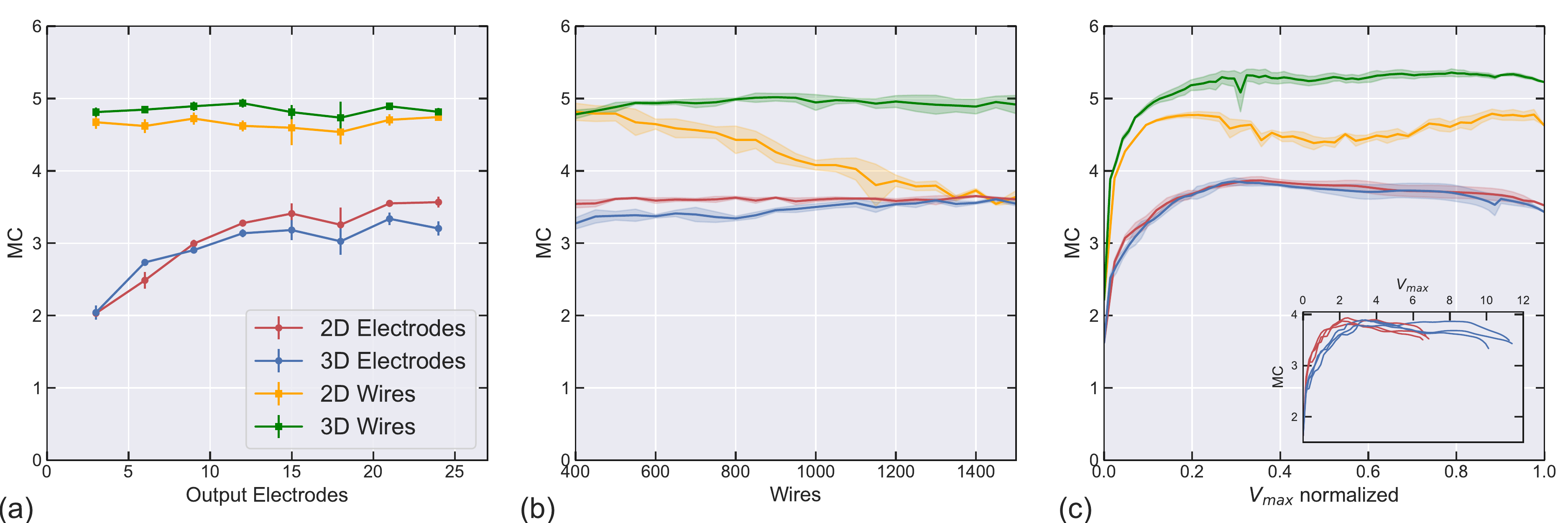}
	\caption{\textit{The memory capacities of the 2D and Q3D networks. (a) An input voltage with $V_{max} = 1$V is fed into a 500NWN. The electrodes are then varied from 3 to 24. The red (2D) and blue (Q3D) lines show the memory capacity when using the readout from output electrodes only. The yellow (2D) and green (Q3D) lines show the memory capacity when every node in the network is used to perform the calculation. (b) With the same input voltage, the electrodes are now fixed at 24, and the number of wires in the networks are varied. We see a constant MC score with low variance when using the electrodes. When the using the readout from every wire, the 2D network MC drops rapidly for increasing wires, while the Q3D remains high. (c) The effect of $V_{max}$ on a 500NWN with 24 electrodes. Although MC remains high for both networks, there is a slight decrease with increasing $V_{max}$. Note that the $x$-scale is the min-max normalized $V_{max}$ value, where the largest voltage is the final voltage the network experiences before one or more junctions move into a high conductance state. Results are averaged across 3 network realizations, and error bars in (a) and shading in (b, c) represent the standard deviation. The inset in (c) shows the results from individual realizations with absolute voltage for the case where the electrode currents are used as predictors.}}
	\label{fgr:memory}
\end{figure*}
Figure \ref{fgr:NLTandMC}a shows a schematic of the MC task. The input to the network, $\mathbf{U}$, is a random sequence of voltage values drawn uniformly from the range $[0, V_{max}]$. The input is applied at the single input electrode and the output signal is collected from the readouts (which can be either the electrodes, as illustrated in Figure \ref{fgr:NLTandMC}a, or every wire). The measured readout sequences are then combined into a predictor matrix, $\mathbf{X}$. $MC_k$ quantifies the performance of the network in reconstructing a  version of the input that is delayed by a time $k$. 
In other words the target $\mathbf{{y}}$ at time $t$ is the same as the input at time $t-k$. The reconstructed signal $\mathbf{\hat{y}}$ (i.e. the prediction of the reservoir) is obtained from a linear combination of the readouts at time $t$. 
For the delay $k$  the required weight vectors, $\mathbf{w}_k = \textbf{X}^{\dagger}\mathbf{y}$, are obtained from the Moore-Penrose Pseudo inverse $\mathbf{X}^{\dagger}$, which minimizes the root mean square error (RMSE)
\begin{equation}
    RMSE = \sqrt{\frac{\sum (\mathbf{y} - \mathbf{\hat{y}})^2}{L}}.
\end{equation}
Here $\mathbf{\hat{y}} = \mathbf{w}_{k}\mathbf{X}$ and $L$ is the length of the input sequence. After training, the  memory capacity for each $k$ is obtained from the predicted output for a different test input sequence, and  is defined as the squared correlation coefficient between the target and  predicted  signals,
\begin{equation}
    MC_k = \frac{cov^2(\mathbf{y}, \mathbf{\hat{y}})}{\sigma^2(\mathbf{y})\sigma^2(\mathbf{\hat{y}})}.
\end{equation}
The total memory capacity is then a sum over all possible delays
\begin{equation}
    MC = \sum_{k=1}^{k_{max}} MC_k.
\end{equation}
\begin{figure*}[t]
	\centering
	\includegraphics[width=\textwidth]{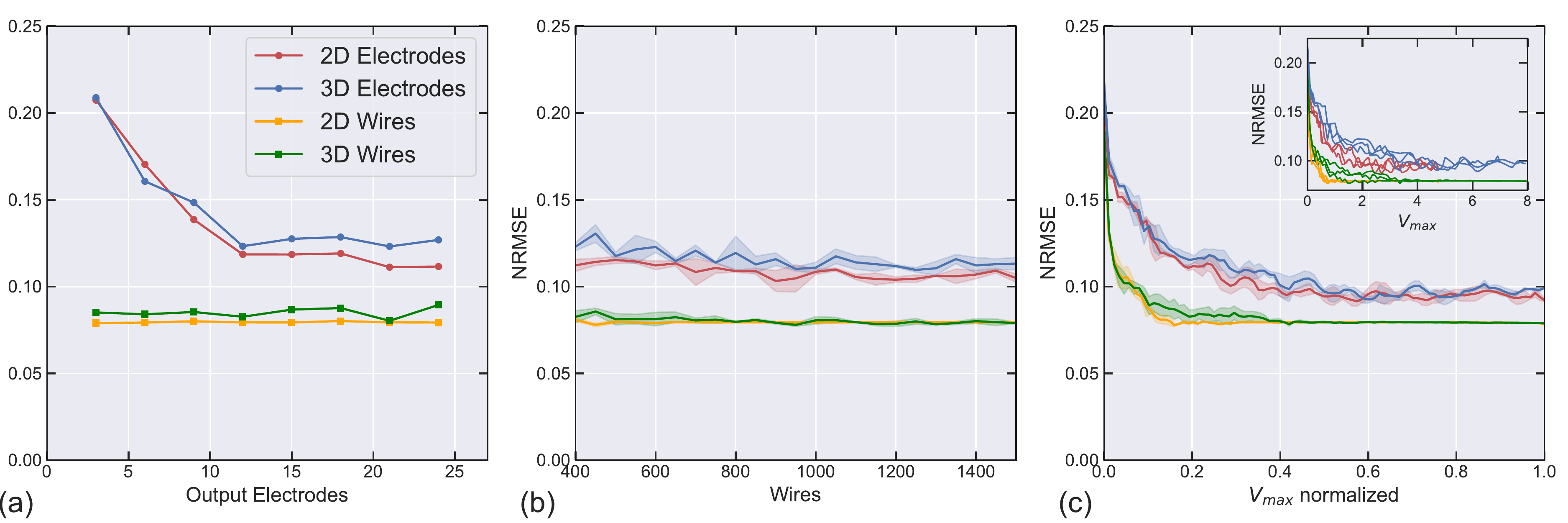}
	\caption{\textit{(a) The effect of increasing the number of electrodes in the NWNs on the non-linear transformation task. Red and blue correspond to the values obtained using the readout from the electrodes for testing in the 2D and Q3D networks respectively; the yellow and green (2D and Q3D respectively) are the results using the readout from every wire in the networks. As the number of electrodes is increased, performance quickly approaches that of the network when the potential from every wire is used. (b) With the same input voltage amplitude (1V), the electrodes are now fixed at 24, and the number of wires in the networks are varied. The NRMSE$_E$ of both networks decrease slightly and plateau after approximately 1000 wires. When using all wires for the task readout, the 2D and Q3D network performance is similar. (c) The effect of increasing the input signal voltage. Note that the $x$-scale is the min-max normalized $V_{max}$ value, where the largest voltage is the final voltage before the network experiences one or more junctions move into a high conductance state. Again, the performance of 2D and Q3D networks are similar. Results are averaged across network realizations, and errorbars in (a) and shading in (b, c) represent the standard deviation. The inset in (c) shows the results from individual realizations with absolute voltage.}}
	\label{fgr:NLTres}
\end{figure*}

\subsection{Nonlinear transformation}
The NLT task demonstrates the ability of the network to perform a non-linear mapping using the higher harmonics generated by the internal dynamics of the network \citep{sillin2013, jaeger2002tutorial}. As illustrated schematically in Figure \ref{fgr:NLTandMC}, an input sine wave $\mathbf{U}$ with signal level between $V_{min}$ and $V_{max}$ is fed into the network at the input electrode. The process of training and testing the network is essentially identical to that of the MC task, except the target function $\mathbf{y}$ is a square wave in the range $[-1,1]$ (Figure \ref{fgr:NLTandMC}b). The normalized root mean square error (NRMSE) between the target function and the transformed function $\mathbf{\hat{y}}$,
\begin{equation}
	NRMSE = \frac{RMSE}{y_{max} - y_{min}}
\end{equation}
is used as the performance metric.

\subsection{Noisy input}
In many machine learning models, adding noise to the input signals acts as a type of regularization \citep{bishop1995} - i.e. it prevents the weights from taking extreme values (overtraining). The result is a model that generalizes better and is more resilient to noisy test data \citep{bishop2006,goodfellow2016}. Therefore, in order to investigate the generalizability of NWN to performing the nonlinear transformation, a degree of random noise is added to the input training signals. The networks are then tested using the regression weights obtained from training.

\section{Results}
The simulations are performed for an experimentally relevant range of parameters, as described in detail in \cite{daniels2021}. For all simulations, the dimensions of the deposition area are kept fixed at $30\times 30 \mu$m$^2$, and the wire lengths are chosen to be $6\mu$m. All input voltage values are positive, which means that the network junctions are operating in a strictly unipolar regime. The values of $\mu$ and $\kappa$ were chosen to produce equal time constants for the growth and decay of the conductive hillock in the low voltage regime. When analyzing the dependence of performance in the MC and NLT tasks upon electrode numbers and $V_{max}$, we used a fixed wire density of $\sim$0.55 NW/$\mu$m$^2$ (500 wires). This allows a direct comparison with recent work in experimental nanowire devices \citep{avizienis2012,diaz2019,milano2020,ocallaghan2018}.

Since a non-singular conductance matrix is required to solve Kirchhoff's laws the number of wires and number of electrodes comprising the network is constrained. As the number of wires is decreased towards the percolation threshold (i.e. the minimum number of wires $N_C$ for which the two sides of the system are fully connected in 50\% of simulations \citep{stauffer2018}), the probability of producing a fully connected network that is connected to all electrodes decreases. We find that for 24 electrodes, arranged as in Figure \ref{fgr:NLTandMC}, a minimum of 400 wires is required to ensure all electrodes are connected. This is consistent with experimental studies where densities of wires are typically well above the percolation threshold \citep{kuncic2021}. 

For each task we considered the impact of the number of output electrodes $E$ ranging from 3 to 24 electrodes, the number of wires $N$ in the network ranging from 400 to 1500 wires, and the value of $V_{max}$ ranging up to to 12V. (Input values higher than 12V consistently result in hillocks forming complete bridges across multiple junctions within any given network.)

\subsection{Memory capacity}
Figure \ref{fgr:memory}a shows the impact of the number of electrodes on the memory capacity in E-mode $MC_E$ and in N-mode $MC_N$. Perhaps surprisingly, given the significant differences in network structure \citep{daniels2021}, $MC_E$ for the 2D (red) and Q3D (blue) networks are strikingly similar. The same is also true of $MC_N$ (yellow for 2D, green for Q3D). In both types of network, as $E$ is increased, $MC_E$ increases gradually but gains in performance for $E>12$ become small. $MC_N$  is not affected by the number of electrodes, and is higher than $MC_E$, consistent with the expectation that readout from all nodes should provide an upper limit on performance. 

In order to determine the impact of $N$ on the memory capacity, we set $E=$ 24 and varied the number of wires. Figure \ref{fgr:memory}b shows that $MC_E$ is always similar for the 2D and Q3D networks (compare red and blue curves) and the difference becomes smaller as $N$ increases. This is surprising, since as $N$ is increased the stacking in the Q3D network becomes more pronounced. $MC_N$ for the 2D NWNs begins to drop off gradually and becomes comparable with that of $MC_E$, whereas $MC_N$ for the Q3D NWNs remains relatively constant. Hence, for large $N$ the upper limit on performance of the Q3D networks is significantly higher than for the 2D networks. As $N$ increases, more wires become shorted together in the 2D networks (decreasing the average degree), and this effectively reduces the number of wires with unique outputs and the results become similar to those obtained from using a small sample of the wires (and the performance using electrode currents). Recent work by \cite{han2021} has also shown that the MC of directed acyclic networks decreases rapidly as the density of nodes in the network increases. These results are also consistent with the findings of \cite{loeffler2021} and \cite{zhu2020b}, which show MC declining with an increase in the mean degree of the networks.

\begin{figure*}[h!]
	\centering
	\includegraphics[width=13cm]{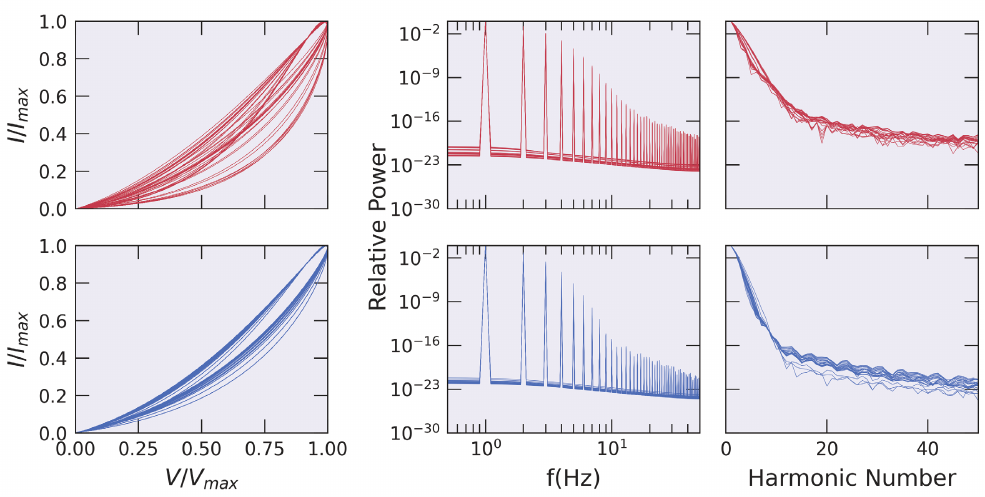}
	\caption{\textit{The nonlinear current response of the output electrodes and the higher harmonics produced from 24-electrode 2D (red) and Q3D (blue) NWNs with a 3V sinusoidal input, showing a highly nonlinear current response and a wide range of higher harmonics.}}
	\label{fgr:higherHarm}
\end{figure*}

Figure \ref{fgr:memory}c shows the impact of the input voltage on the memory capacity for a fixed $E=$ 24. The 2D and Q3D networks were subjected to the same voltage input sequence, but for each realization of the networks, the range of $V_{max}$ was constrained by the requirement that all junctions remain in the low voltage regime (i.e. that no hillock formed a complete bridge across any junction). Figure \ref{fgr:memory}c shows the change in MC with $V_{max}$ (the values of $V_{max}$ on the $x$-axis are min-max normalized which allows us to compare performance relative to the critical voltage, $V_C$). Initially, an increase in input voltage causes a rapid increase in both $MC_N$ and $MC_E$, as a greater degree of nonlinearity in the junctions can be exploited. However, as $V_{max}$ increases further there is a saturation and then a slight decline. This effect is also seen in directed acyclic networks: as the input signal is increased, MC decreases \citep{han2021}.

The inset of Figure \ref{fgr:memory}c shows $MC_E$ of each realization of the system as a function of the absolute voltage. Nearly twice the input voltage can be applied to the Q3D networks compared to the 2D networks. This is consistent with previous work \citep{daniels2021} showing that the Q3D networks have longer path lengths -- as voltage is distributed across many junctions in series, the longer path between electrodes ensures a lower voltage across each junction.

\subsection{Nonlinear transformation}
Figure \ref{fgr:NLTres} compares the performance in the NLT for different network types. The input sequence used was a 1V sinusoidal signal of 1Hz frequency for 21 periods. The network contains $N=500$ nanowires. In Figure \ref{fgr:NLTres}a the red and blue curves show the performance in E-mode (NRMSE$_E$) for the 2D and Q3D networks respectively. As $E$ is increased, NRMSE$_E$ drops (i.e. performance improves), with the 2D networks slightly outperforming the Q3D networks.  Figure \ref{fgr:NLTres} also shows the theoretical maximum performance, NRMSE$_N$, of the networks  in $N$-mode for the 2D and Q3D networks (yellow and green respectively). NRMSE$_E$  approaches NRMSE$_N$ as $E$ is increased, and for $E>12$ there is very little performance gain, as was also observed for the MC task. When using N-mode to perform the NLT, the testing and training NRMSE$_N$ are similar to one another for the Q3D networks, yet the testing NRMSE$_N$ is significantly higher than the training NRMSE$_N$, suggesting that the 2D network weights are overfitting the training data (not shown).

Figure \ref{fgr:NLTres}b shows the impact of $N$ on the nonlinear tranformation performance. As with the MC task, $E$ is fixed at 24, and $N$ ranges from 400 to 1500. In E-mode, performance improves slightly with increasing $N$, until $N\sim$1000. When using N-mode, performance remains constant for $N>500$ wires, with very little variance across different realizations in both networks.
\begin{figure*}[t]
	\centering
	\includegraphics[width=\textwidth]{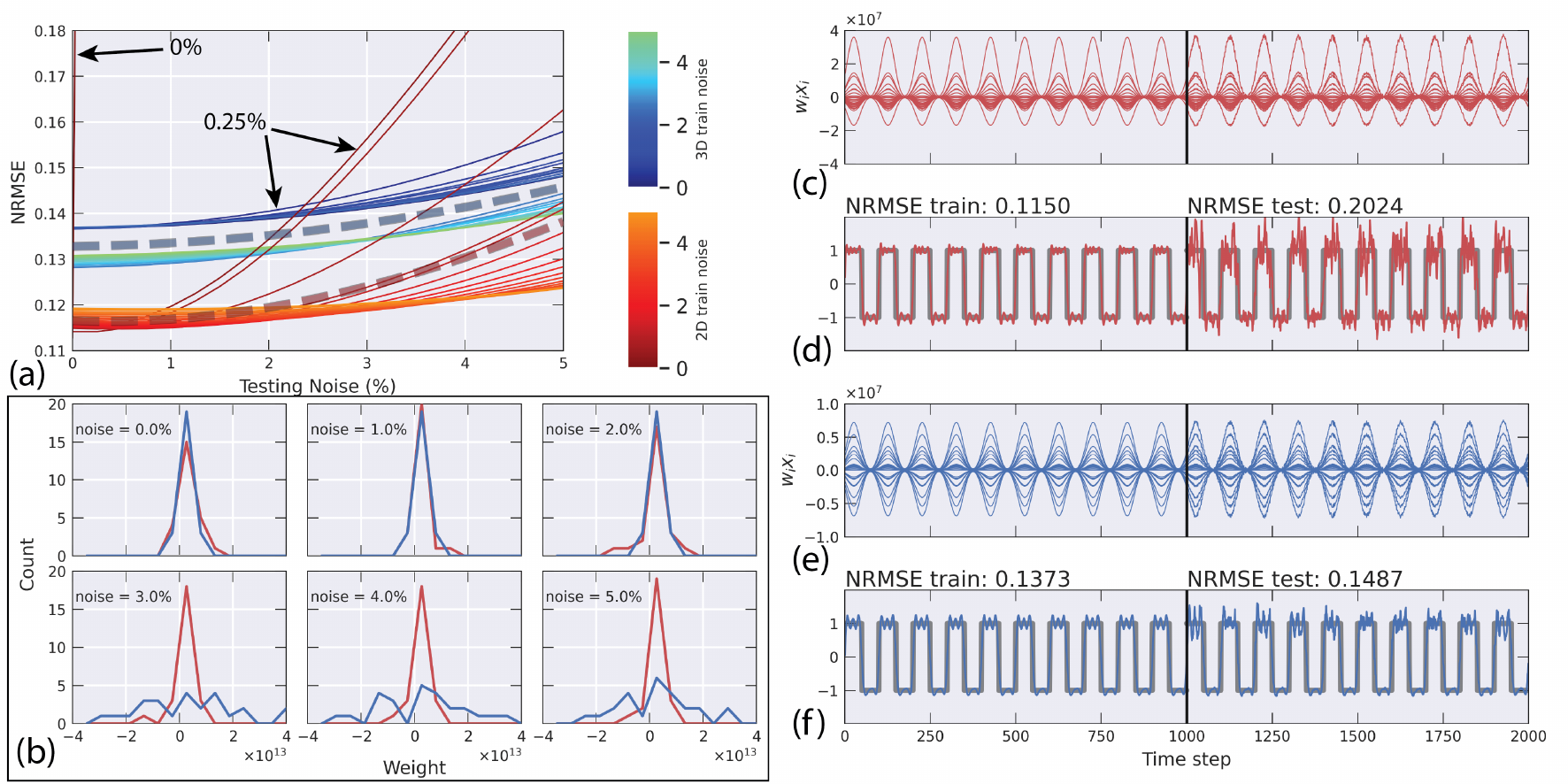}
	\caption{\textit{Comparison of performance in the presence of noise. (a) When the networks are trained on low levels of noise, the 2D networks show a high sensitivity to subsequent noisy test inputs. Despite initially performing worse than the 2D networke much more resilient and the 2D network consistently outperforms the Q3D. Note the 2D network trained on 0\%, marked by the arrow. The bold dotted lines are the mean NRMSE scores for the two different networks. (b) The weights of the 2D and Q3D networks (red and blue) trained on a range of noise. Shown here is a sample of the networks trained on integer amounts of noise. (c-f) An example of the Q3D network (blue) resilience to low amounts of training noise compared with the 2D networks (red). The example networks were trained on 0.25\% noise and then tested on 5\%. Shown in (c) and (e) are the weighted outputs from the networks. The corresponding noise curves are marked in (a) by the two arrows.}}
	\label{fgr:noise}
\end{figure*}

Figure \ref{fgr:NLTres}c shows that as $V_{max}$ increases, there is a gradual improvement in performance, which then plateaus at 3-4V for both networks. There is very little difference in performance between the 2D and Q3D networks. This is once again surprising given the different topological structure of the 2D and Q3D networks. However, as shown in the inset of Figure \ref{fgr:NLTres}c, the Q3D network can be subjected to nearly twice the input amplitude compared to the 2D networks, achieving consistent NRMSE scores over a wider range of voltage inputs.

In order to successfully transform the input signal into the target, the network must be capable of higher harmonic generation (HHG). We therefore also examined the higher harmonics produced by each network.

Figure \ref{fgr:higherHarm} shows the HHG for an example 2D and an example Q3D network both in a 24 electrode configuration, showing a highly nonlinear network response to a sinusoidal input, and the generation of a wide range of higher harmomics. The generation of the higher harmonics can be largely attributed to the nonlinear response of the network. Hence the network is able to perform a nonlinear mapping of the input features to a higher-dimensional space. Note that in Figure \ref{fgr:higherHarm}, we can see that the Q3D network has a smaller degree of hysteresis. This is likely due to the lower voltage across each junction caused by the shorter mean path lengths within the Q3D network.

\subsection{Noisy input}
In order to test the resilience of the network performance, we trained the output weights for each of the networks in the same manner as for the NLT task, but added input noise ranging from 0 to 5\% of the input signal amplitude. We then exposed the trained networks to testing inputs with different amounts of noise.

When the networks receive training input with small amounts of noise ($<1\%$), the Q3D networks achieve a lower NRMSE on the testing signal than the 2D networks. In fact, when the networks are trained in the absence of noise, the test performance for the 2D network is poor when even the slightest amount of noise is added to the test input (see the nearly vertical red line close to the vertical axis of Figure \ref{fgr:noise}a). When the 2D networks are trained on input signals with greater amounts of noise, the testing performance significantly improves.

In general, the 2D networks have a slightly better performance (lower NRMSE) than the Q3D networks. However, the Q3D networks show a significant jump in performance when trained on input with noise above $\sim$2\%. The average performance (bold dashed lines) of the two networks begins to converge for higer levels of noise. Figure \ref{fgr:noise}b shows the evolution of the weights as more input noise is added. When trained with low noise, the distribution of weights in both networks is narrow.
As the noise on the training input is increased, the weight distribution for the 2D networks remains stable, but the  weights for the Q3D network undergo a sharp transition and become much more widely distributed.

Figure \ref{fgr:noise}(c-f) show examples of NLT performance for a 2D and Q3D network exposed to a low amount of input noise (0.25\%). The 2D network achieves a better training performance, but the Q3D network achieves a significantly better test NRMSE. The high values of the weights when trained on low noise and the comparatively poor performance of the networks on the testing data suggest that the regression of the output from the 2D network is overfitting. Adding noise acts to regularize the output weights, allowing it to generalize better on the more noisy test data, as seen by the flatter curves in Figure \ref{fgr:noise}a for the Q3D networks. The slightly narrower range of weights for the Q3D network trained on low noise avoids over fitting and therefore makes them more resilient.

\section{Conclusion}
We have compared the performance of 2D and Q3D nanowire networks in reservoir computing tasks using realistic simulations of the physical systems. The networks have a strikingly similar performance, which is surprising given that the networks have very different topologies. The memory capacities of both networks when using the physically realistic output electrodes are largely unaffected by the key parameters: electrode number, wire number, and voltage input. The  upper bounds on the memory capacity, obtained using the (physically inaccessible) potentials of every wire, are higher for the Q3D networks, especially at large $N$. Our results show that networks with an experimentally achievable number of electrodes are -- perhaps surprisingly -- close to the upper bounds. 

For the nonlinear transformation task, there are again a number of similarities in the performance of the 2D and Q3D networks. The two types of network achieve similar NRMSE scores, supported by the fact that both networks produce a similar range of higher harmonics. The performance when using the electrode outputs is again close to the maximum achievable performance (using the voltage from every wire). For both RC tasks, there is almost no gain in performance for $E>$ 12 (for systems of fixed size and wire density) and so 12 output electrodes appear to be sufficient for practical devices.

There are, however, also some important differences between the network types. The most striking performance difference is the rapid MC performance dropoff for the 2D network when using every wire as a readout. This could be due to a combination of factors, including a shorting together of an increasing number of wires, and a change in the mean degree and mean path lengths across the network. This leads to a decrease in the number of independent nodes and hence an effective decrease in the richness of network outputs. The Q3D network overall is much more robust to changes in the input parameters. When trained with no noise, the 2D networks do a poor job of generalizing to noisy data compared with the Q3D networks. As the 2D networks are trained on signals with more noise, performance decreases slightly but the generalizability improves.

We emphasize that the focus of this work has been to compare performance of the 2D and Q3D networks for identical parameters. However, the question naturally arises as to whether it might be possible to vary the parameters in order to achieve the same network characteristics for the two different dimensionalities. In fact it has already been shown that 3D networks with $N>150$ have distinctly different network characteristics to all 2D networks (see Fig. 6 in \cite{daniels2021}), and so this is not possible.

Given previous literature which emphasizes the importance of network topology on RC performance, it is very surprising that there are not more radical differences in performance between the 2D and Q3D networks. This could in part be due to the level of complexity of the tasks - it is possible that there could be larger differences in performance for more complex computing tasks, and these should be explored in future work. Finally, it is interesting to note that recent investigations of RC performance using networks based on the human connectome reflect the complex interplay between the topology and dynamics of the network \citep{suarez2021}; detailed comparisons of performance for different systems are still required to understand the optimal network characteristics for different tasks.

\section*{Declaration of competing interest}
The authors declare that they have no known competing financial interests or personal relationships that could have appeared to influence the work reported in this paper.

\section*{Acknowledgements}
We would like to thank Kourosh Neshatian and James Atlas for useful discussions and input. This project was financially supported by The MacDiarmid Institute for Advanced Materials and Nanotechnology, the Ministry of Business Innovation and Employment, and the Marsden Fund.


\bibliographystyle{elsarticle-harv} 
\bibliography{RCrefs}





\end{document}